\documentclass[12pt]{article}
\usepackage{epsfig,a4}

 \advance\oddsidemargin by -1in
 \advance\evensidemargin
 by -1in
 \marginparsep
 0.4cm
 \advance\topmargin by
 -0.0in
 \makeindex

 \newcommand\la{\langle}
 \newcommand\ra{\rangle}
 \newcommand\beq{\begin{equation}}
 
 \newcommand\eeq{\end{equation}}                                               
 \newcommand\beqn{\begin{eqnarray}}
 \newcommand\eeqn{\end{eqnarray}}

\def\la{\langle}
\def\ra{\rangle}

\def\GeV{\,\mbox{GeV}}
\def\eps{\varepsilon}

\def\lsim{\mathrel{\rlap{\lower4pt\hbox{\hskip1pt$\sim$}}
\raise1pt\hbox{$<$}}}
\begin{document} 
 
\title{\bf Long-Range Coulomb Forces in DIS:\\ Missed Radiative 
Corrections?}

\maketitle
 
\begin{center}
 
 {\large B.Z.~Kopeliovich$^{1-3}$, A.V.~Tarasov$^{1-3}$ and 
O.O.~Voskresenskaya$^{3}$}
 \\[1cm]
$^{1}${\sl Max-Planck Institut f\"ur Kernphysik, Postfach 103980, 69029
Heidelberg, Germany}\\[0.2cm]
$^{2}${\sl Institut f\"ur Theoretische Physik der Universit\"at, 93040
Regensburg, Germany} \\[0.2cm]
$^{3}${\sl Joint Institute for Nuclear Research, Dubna, 141980 Moscow
Region, Russia}\\[0.2cm]
 
\end{center}

\vspace{1cm}
 
\begin{abstract}

The Born approximation, one photon exchange, used for DIS is subject to virtual
radiative corrections which are related to the long-range Coulomb forces. They may
be sizeable for heavy nuclei since $Z\alpha$ is not a small parameter. So far
these corrections are known only for two processes, elastic scattering and
bremsstrahlung on the Coulomb field of a point-like target. While the former
amplitude acquires only a phase, in the latter case the cross section is modified
also. Although the problem of Coulomb corrections for DIS on nuclei is extremely
difficult, it should be challenged rather than 'swept under the carpet'. The
importance of these radiative corrections is questioned in present paper. We show
that in the simplest case of a constant hadronic current the Coulomb corrections
provide a phase to the Born amplitude, therefore the cross section remains the
same.  Inclusion of more realistic hadronic dynamics changes this conclusion. The
example of coherent production of vector mesons off nuclei reveals large effects.
So far a little progress has been made deriving lepton wave functions in the
Coulomb field of an extended target. Employing available results based on the
first-order approximation in $Z\alpha$, we conclude that the Coulomb corrections
are still important for heavy nuclei. We also consider an alternative approach for
extended nuclear targets, the eikonal approximation, which we demonstrate to
reproduce the known exact results for Coulomb corrections. Calculating
electroproduction of vector mesons we again arrive at a large deviation from the
Born approximation. We conclude that one should accept with caution the
experimental results for nuclear effects in DIS based on analyses done in the Born
approximation.

\end{abstract}
 
 
\newpage
 
\section{Introduction}

\begin{figure}[tbh]   
\includegraphics{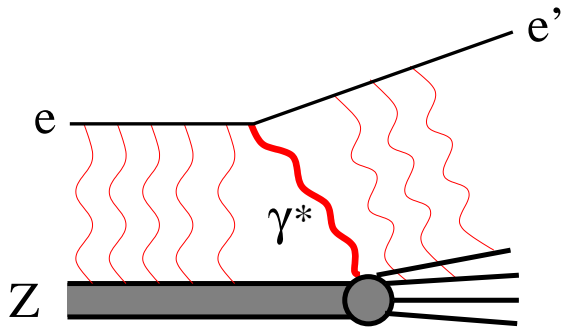}                                    
\begin{center}        
\vspace{5cm}  
\parbox{13cm}
 {\caption[Delta]
 {\sl Deep-inelastic electron-nucleus scattering.
The thin wavy lines illustrate the long-range Coulomb
forces which violate the DIS kinematics 
($\vec q\not= \vec p_1-\vec p_2$) and modify the 
lepton wave functions.}
\label{dis}}
\end{center}
\end{figure}

Smallness of the fine structure constant $\alpha$ usually justifies
lowest order perturbative QED calculations. In some cases, however,
the expansion parameter is not small, for instance, for interactions
with heavy nuclei, where $Z\alpha \sim 1$. Therefore, the validity of 
Born approximation for deep-inelastic scattering (DIS) should be questioned
since the incoming and outgoing leptons propagate in the Coulomb field
of the target, as is illustrated in Fig.~\ref{dis}, which can cause a deviation
of the DIS cross section from the Born form.  
This may lead to important modifications of experimental data
for nuclear effects in DIS which rely on analyses based on the Born 
approximation.

The simplest example is the elastic lepton scattering in the Coulomb
field of a nucleus. Neglecting the nuclear structure, i.e. treating the
target as a point-like charge, one can solve this problem exactly. Gordon
\cite{g} and Mott \cite{mott} have done it in the framework of
nonrelativistic quantum mechanics. It turns out that the resulting
amplitude is different from the Born approximation only by a phase which
has opposite signs for positive and negative leptons. Thus, the elastic
cross section has no Coulomb corrections.  On the other hand, in the case
of hadronic elastic scattering the Coulomb effects do modify the cross
section because of the relative Coulomb phase \cite{bethe,cahn,kt}.

In the relativistic case the problem of elastic Coulomb scattering was
solved for spinless particles by Schr\"odinger \cite{schr} and for
fermions by Mott \cite{mott} and Darwin \cite{darwin}. Although the
scattering phases were calculated, summation of the partial amplitudes,
i.e. construction of the wave function of a relativistic charged particle
in the Coulomb field, is still a challenge.

Nevertheless, for practical applications one can employ the approximation
of high orbital momenta, $L^2 \gg (Z\alpha)^2$, which helps to solve
the problem. Corresponding wave functions in the Coulomb field were found
by Furry \cite{furry} and Maue and Sommerfeld \cite{sommerfeld}.

Bethe and Maximon \cite{maximon} made use of the same approximation to
solve a more complicated problem of bremsstrahlung and production of
electron-positron pairs by an electron scattering in the Coulomb field.
Their finding is interesting, the long-range Coulomb forces modify not
only the phase, but lead to a suppression of the cross section compared
to the Born approximation. The effect is significant, about $10 - 15\,\%$
for heavy nuclei. The fact that the Coulomb forces suppress radiation can
be intuitively understood as a manifestation of the Landau-Pomeranchuk
effect. Indeed, the electron trajectory is bending smoothly over long
distances, and the photons radiated from different parts of the electron
path interfere destructively.

One may expect similar modifications in other reactions, especially those
which involve strong interactions, like DIS, they should be revisited
aiming to clarify the role of long-range Coulomb forces.  All analyses of
DIS data are based on the Born approximation for the electromagnetic
interaction with the target, i.e. the wave functions of the upcoming and
outgoing leptons are treated as plane waves. At the same time, precise
measurements \cite{nmc} disclose rather fine, few percent nuclear
effects, like small antishadowing at $x\sim 0.1$, or variation of nuclear
shadowing with $Q^2$.  It is still questionable how these small effects
are related to the missed distortions generated by the Coulomb field.

On the other hand, the HERMES experiment \cite{hermes} has discovered
recently unusually strong nuclear effects which cannot be understood
within conventional approaches \cite{krt2}. Since these effects arise
only after the radiative corrections are introduced into the analysis, it
also motivated us to have a closer look at the problem of reliability of
radiative corrections.

This paper is organized as follows.  In Sect.~\ref{theorem} we start with
a specific simplified example of DIS with a hadronic current independent
of $Q^2$. In this case the amplitude is proven to gain only a phase
leaving the cross section unmodified. This is a new nontrivial result
which we utilize for further applications in this paper.

As a more realistic, but still simple example we choose the coherent
vector meson production off nuclei which is treated within the vector
dominance model (VDM). In Sect.~\ref{point-like} we consider the case of
a point-like target and arrive at a substantially modified cross section
which deviates from the Born approximation.  The Coulomb corrections
increase with $Q^2$ as is shown in Fig.~\ref{vdm}.

The next step toward a realistic hadronic current is done 
in Sect.~\ref{slope}. It corresponds to the vector meson production off 
an extended nucleus, but with the same leptonic wave functions
calculated for a point-like Coulomb center.
Again large corrections to the cross section are found whose value and sign 
varies with lepton energy and $Q^2$, as is demonstrated in Fig.~\ref{vdm1}.

Eventually, in Sect.~\ref{large-nucl} we replace the exact wave functions
of the lepton in the Coulomb field of a point-like charge by approximate
ones calculated for the case of an extended nuclear target. Unfortunately
the only solution available in the literature corresponds to the first
order expansion in $Z\alpha$. Assuming that it is not a phase we arrive
at a significant correction up to $13\%$ for heavy nuclei.

In Sect.~\ref{eikonal} try a different technique, the eikonal
approximation, which is designed in a way which allows to perform
calculations for nuclei of large size. First of all, we check whether the
eikonal approximation is able to reproduce the Coulomb corrections
previously calculated exactly. We demonstrate in \ref{b} for the case of
$e^+e^-$ production off nuclei that the exact results for the Coulomb
corrections are fully reproduced by the eikonal approximation. We
calculate the Coulomb corrections for the process of vector meson
production in the eikonal approximation in the limit of very high
energies. Again the corrections are found to be large and to saturate as
function of $Q^2$ at $Q^2>0.1\,GeV^2$, as is shown in
Fig.~\ref{eikonal-fig}.

Although our results are too rough to be incorporated into analyses of
data for DIS, we conclude in Sect.~\ref{conclusions} that one should
accept with a precaution the available experimental results for nuclear
effect in DIS. We have found only one specific case of an oversimplified
hadronic current when the Coulomb corrections have the form of a phase.
In more realistic situations the hadronic current consists of a few (or
many) terms which acquire different phases and lead to a modified cross
section. Thus, all available data for DIS on heavy nuclei based on
analyses utilizing the Born approximation might change if the Coulomb
corrections are applied.

\section{A little theorem}\label{theorem}

This section is aimed to demonstrate that for a specific simple form of
the conserved hadronic current the effects of long-range Coulomb forces
are reduced to a phase factor, like it happens for elastic scattering.
This result then will be implemented into more complicated situations.

Let us start with a process which leaves the nucleus intact, for example
coherent electroproduction of vector mesons. The amplitude of this 
reaction in Born approximation reads,
 \beq
M(e\,A \to e'\,V\,A) \propto
\int d^4x\,d^4y\,\la p_2|j_\mu(x)|p_1\ra\,
D_{\mu\nu}(x-y)\,\la V|J_\mu(y)|0\ra\ ,
\label{10}
 \eeq
 where 
 \beq
\la p_2|j_\mu(x)|p_1\ra = \bar\Psi_{p_2}(x)\,
\gamma_\mu\,\Psi_{p_1}(x)
\label{20}
 \eeq
 is the operator of lepton current; $J_\mu(y)$ is the hadronic current
operator; $D_{\mu\nu}(x-y)$ is the Green function of the photon. Using
the Feynman gauge we have,
 \beq 
D_{\mu\nu}(x-y) = -g_{\mu\nu}\, 
\int d^4Q\, \frac{e^{iQ(x-y)}}
{Q^2+io}\ , 
\label{30}
 \eeq 
and
 \beq
M(e\,A \to e'\,V\,A) \propto
-\int \frac{d^4Q}{Q^2+io}\,
j_\mu(Q;p_1,p_2)\,J_\mu(Q;P_V)\ .
\label{40}
 \eeq 
 Here the Fourier transform of the currents reads,
 \beqn
j_\mu(Q;p_1,p_2) &=& \int d^4x\,
\la p_2|j_\mu(x)|p_1\ra\,e^{iQx}\ ;
\\
J_\mu(Q;P_V) &=& \int d^4y
\,\la P_V|J_\mu(y)|0\ra\,e^{-iQy}\ ;
\label{50}
 \eeqn
$p_{1,2}$ are the initial and final lepton momenta; $P_V$ is the
momentum of the produced $V$.

If the initial and final wave functions of the lepton are undistorted
plane waves,
 \beq
\Psi_{p_1,p_2}(x) = e^{ip_{1,2}x}\,u(p_{1,2})\ ,
\label{60}
 \eeq
then 
 \beq
j_\mu(Q;p_1,p_2) = (2\pi)^4\,\delta(p_1-p_2-Q)\,
\bar u(p_2)\gamma_\mu u(p_1)\ ,
\label{70}
 \eeq
 and
 \beq
M(e\,A \to e'\,V\,A) \propto
\frac{1}{(p_1-p_2)^2}\,
\bar u(p_2)\gamma_\mu u(p_1)\,
J_\mu(p_2-p_1;P_V)\ .
\label{80}
 \eeq

However, in the presence of a Coulomb field one must rely on the solution
of the Dirac equation, rather than on the plane waves,
 \beq
i\,\frac{\partial}{\partial\,t}\,\Psi(\vec r,t) = 
\left[- i\,\vec\alpha\cdot\vec\nabla +
\beta\,m + e\,\phi(r)\right]\,\Psi(\vec r,t)\ .
\label{90}
 \eeq
 Here $\vec\alpha$ and $\beta$ are the standard Dirac matrixes;  
$\phi(r)$ is the Coulomb potential which is independent of energy in the
rest frame of the nucleus; $m$ is the lepton mass. The initial state
leptonic wave function which is a solution of this equation, has the form
of sum of a plane and a spherical outgoing waves, $\Psi^+(\vec p_1,\vec
r)$ at $\vec r \to \infty$, while the final state wave function should
contain a plane and an incoming waves, $\Psi^-(\vec p_2,\vec r)$.

In general case equation (\ref{90}) can be solved only numerically, 
but if  the effects of nuclear size can be neglected, i.e. 
 \beq 
\phi(r) = - \frac{Ze}{r}\ ,
\label{100}
 \eeq
 an approximate analytical solution exists, as it was found by W.H.~Furry
\cite{furry},
 \beqn
\Psi^+(\vec p_1,\vec r) &=&
\frac{C}{\sqrt{2\epsilon_1}}\,
e^{i\vec p_1\cdot\vec r}\,
\left(1-\frac{i\vec\alpha\cdot\vec\nabla}
{2\epsilon_1}\right)\,
F\left[iZ\alpha,1;i(p_1r-\vec p_1\cdot\vec r)\right]
\ u(\vec p_1)\ ,\label{105}\\
\Psi^-(\vec p_2,\vec r) &=&
\frac{C^*}{\sqrt{2\epsilon_2}}\,
e^{i\vec p_2\cdot\vec r}\,
\left(1-\frac{i\vec\alpha\cdot\vec\nabla}
{2\epsilon_2}\right)\,
F\left[-iZ\alpha,1;-i(p_2r+\vec p_2\cdot\vec r)\right]
\ u(\vec p_2)\ ,
\label{110}
 \eeqn
 where
 \beqn
C &=& e^{\pi Z\alpha/2}\,\Gamma(1-iZ\alpha)\ ,
\nonumber\\
\epsilon_{1,2} &=& \sqrt{\vec p_{1,2}^{\,2} +m^2}\ ,
\label{120}
 \eeqn
 $u(\vec p_{1,2})$ are the 4-spinors of the leptons, $\Gamma(x)$ is the
gamma-function, and $F(a,b;c)$ is the confluent hypergeometric function.
The condition that the leptons are ultrarelativistic, $\epsilon_{1,2}\gg
m$, implies that the essential orbital momenta are large, $l\gg1$. It is
well satisfied in all cases we are interested in.

For the sake of simplicity we will work with ``spinless leptons''
since we are interested only in an estimate of the effects and do not expect
a principal modification related to the lepton spin.
Then the lepton wave functions satisfy the Klein-Gordon equation,
 \beqn
- \Delta\,\Psi(r) &=& 
\left\{\left[\epsilon - e\,\phi(r)\right]^2 -
m^2\right\}\,\Psi(r)\nonumber\\
&=& \left[p^2-2\,e\,\epsilon\,\phi(r) + e^2\,\phi^2(r)
\right]\,\Psi(r)\ .
\label{130}
 \eeqn Apparently, at high energies the term $e^2\,\phi^2(r)$ can be
neglected, and for the Coulomb potential Eq.(\ref {100}) one can get the
exact solution,
 \beqn
\Psi^+(\vec p_1,\vec r) &=&
C\,F\left[iZ\alpha,1;i(p_1r-\vec p_1\cdot\vec r)\right]
e^{i\vec p_1\cdot\vec r}\,
\ ,\label{140}\\
\Psi^-(\vec p_2,\vec r) &=&
C^*\,F\left[-iZ\alpha,1;-i(p_2r+\vec p_2\cdot\vec r)\right]\,
e^{i\vec p_2\cdot\vec r}\ ,
\label{150}
 \eeqn
 Correspondingly, for the lepton current,
$j_\mu(x) = j_\mu(\vec r,t)$,
 \beq
j_\mu(\vec r,t) = e^{i(\epsilon_1-\epsilon_2)t}\,
j_\mu(\vec r)\ ,
\label{155}
 \eeq
where
 \beqn
j_0(\vec r) &=& e\,
\Psi^-\Bigr.^*(\vec p_2,\vec r)\,
\left(\epsilon_1+\epsilon_2+
\frac{2Z\alpha}{r}\right)\,
\Psi^+(\vec p_1,\vec r)\ ,
\label{160}\\
\vec j(\vec r) &=& 
i\,e\,\left[\vec\nabla\,
\Psi^-\Bigr.^*(\vec p_2,\vec r)\,
\Psi^+(\vec p_1,\vec r) - 
\Psi^-\Bigr.^*(\vec p_2,\vec r)\,
\vec\nabla\,
\Psi^+(\vec p_1,\vec r)\right]\ .
\label{170}
 \eeqn
This current is conserved,
 \beq
\frac{\partial j_0(\vec r,t)}
{\partial t} = 
\vec\nabla\cdot\vec j(\vec r,t)\,
\label{180}
 \eeq
 or, in momentum representation,
 \beq
\nu\,j_0(Q) = \vec q\cdot\vec j(\vec q)\ ,
\label{190}
 \eeq
 where $\nu$, $\vec q$ are the energy and momentum transferred to the
target, $\nu^2-\vec q^{\,2}=-Q^2$, and Fourier transform of the current
reads,
 \beq
j_\mu(Q)= \int d^4x\,e^{iQx}\,j_\mu(x)\ .
\label{200}
 \eeq
In terms of time-independent current it reads,
 \beqn
j_0(Q) &=& 2\pi\delta(\epsilon_1-\epsilon_2-\nu)
\int d^3r\,e^{i\vec q\cdot\vec r}\,
j_0(\vec r)\ ;
\label{210}\\ 
\vec j(Q) &=& 2\pi\delta(\epsilon_1-\epsilon_2-\nu)
\int d^3r\,e^{i\vec q\cdot\vec r}\,
\vec j(\vec r)\ .
\label{220}
 \eeqn
Using these expressions and conservation of the hadronic 
current,
 \beq
\nu\,J_0(Q) = \vec q\cdot\vec J(Q)\ ,
\label{230}
 \eeq
we arrive at the following form of the amplitude Eq.~(\ref{40}),
 \beq
M = 2\pi\int\frac{d^3q}{-Q^2-io}\,
\left[\vec j(\vec q) - j_0(\vec q)\,
\frac{\vec q}{\nu}\right]\,\vec J(Q)\ .
\label{240}
 \eeq 
 
The integral in Eq.~(\ref{240}) contains hadronic current 
$\vec J(Q)$ with unknown dependence on $Q$ [$Q$ dependence of the 
leptonic current is fixed by Eqs.~(\ref{210}) - (\ref{220})]. As the first 
simple trial we assume that it is $Q$ independent, $\vec J(Q)=\vec c$.
Then the amplitude Eq.~(\ref{240}) takes the form,
 \beq
M = 2i\,(2\pi)^2\,\frac{\vec c\cdot\vec d}{\nu}\ ,
\label{250}
 \eeq
 where 
 \beq
\vec d = {1\over2}\int d^3r\,\frac{e^{i\nu r}}{r}\,
\left[i\,\vec\nabla\,j_0(\vec r) + \nu\,\vec j(\vec r)\right]\ .
\label{260}
 \eeq

Let us consider kinematics when the virtual photon takes a finite
fraction $y=\nu/\epsilon_1$ of the initial lepton energy, and the
scattering angle is small, $\theta\ll 1$. In this case the relative
contribution of the the term $Z\alpha/r$ in the current Eq.~(\ref{160})
estimated with the plane wave approximation is small,
$\ln(1/\theta)\,\theta^2/y \ll 1$. Thus, it is suppressed by a factor
$x_{Bj}\,m_N/\nu$. Neglecting this term we simplify the current
Eq.~(\ref{160}) to
 \beq
j_0(r) = (\eps_1+\eps_2)\,
\Psi^-\Bigr.^*(\vec p_2,\vec r)\,
\Psi^+(\vec p_1,\vec r)\ .
\label{270}
 \eeq
Applying Eqs.~(\ref{140}) - (\ref{150}) to Eqs.~(\ref{170})  - 
(\ref{270}) 
we get for the vector $\vec d$ Eq.~(\ref{260}),
 \beq
\vec d = \int d^3r\,\frac{
e^{i\nu r+i\vec q\cdot\vec r}}{r}\,
\left[\left(\epsilon_1\vec p_2-\epsilon_2\vec p_1\right)\,
F_1(\vec r)\,F_2(\vec r)\, +
i\,\epsilon_1\,F_1(\vec r)\,\vec\nabla
F_2(\vec r) + i\,\epsilon_2\,\vec\nabla F_1(\vec r)\,
F_2(\vec r)\right]\ ,
\label{280}
 \eeq
 where
 \beqn
F_1(\vec r,\vec p_1) &=& C\,F[iZ\alpha,1;i(p_1r - \vec p_1\cdot\vec r)]
\ ,\nonumber\\
F_2(\vec r,\vec p_2) &=& C^*\,
F[-iZ\alpha,1;-i(p_2r + \vec p_2\cdot\vec r)]\ .
\label{285}
  \eeqn

Then we make use of the following relations,
 \beq
\vec\nabla_{\vec r}F_1(\vec r,\vec p_1) = 
-\frac{p_1}{r}\,\vec\nabla_{\vec p_1}
F_1(\vec r,\vec p_1);
\label{290}
 \eeq
 \beq
\vec\nabla_{\vec r}F_2(\vec r,\vec p_2) = 
\frac{p_2}{r}\,\vec\nabla_{\vec p_2}
F_2(\vec r,\vec p_2);
\label{300}
 \eeq
 \beq
{1\over r} = \int\limits_{0}^{\infty} 
d\lambda\,e^{-\lambda r}\ ;
\label{310} 
 \eeq
and (see in \cite{nordsieck})
 \beqn
&& \int \frac{d^3r}{r}\,
F_1(\vec r,\vec p_1)\,F_2(\vec r,\vec p_2)\,
\exp(i\vec q\cdot\vec r - \lambda r + i\nu r)
\equiv I(\vec q,\vec p_1,\vec p_2,\lambda)\nonumber\\
&=& 
\frac{4\pi N}{w}\,\left(\frac{w}{z}\right)^{iZ\alpha}\,
F(iZ\alpha,1-iZ\alpha;1;x)\ ,
\label{320}
 \eeqn
where $F(a,b;c;d)$ is the conventional hypergeometric function,
 \beqn
x &=& 1-\frac{uv}{wz}\ ;\nonumber\\
w &=& \vec q^{\,2} + 
(\lambda - i\nu)^2\ ;\nonumber\\
u &=& (\vec p_2+\vec q)^2 - 
(p_2+\nu+i\lambda)^2\ ;\nonumber\\
v &=& -(\vec p_1-\vec q)^2 +
(p_1+\nu+i\lambda)^2\ ;\nonumber\\
z &=& (p_1+p_2+\nu+i\lambda)^2 -
(\vec p_1-\vec p_2-\vec q)^2\ ;\nonumber\\
N &=& \Bigl|\Gamma(1-iZ\alpha)\Bigr|^2 = 
\frac{\pi Z\alpha}{\sinh(\pi Z\alpha)}\ .
\label{330}
 \eeqn
With Eqs.~(\ref{290}) - (\ref{330}) we arrive at a new expression 
for $\vec d$ in Eq.~(\ref{280}),
 \beq
\vec d = (\epsilon_1\vec p_2 - \epsilon_2\vec p_1)\,
I(\vec q,\vec p_1,\vec p_2,\lambda=0)
\Bigr|_{\vec q=\vec p_1-\vec p_2} -
\int\limits_0^\infty d\lambda\,
\vec g(\vec p_1,\vec p_2,\lambda)\ ,
\label{340}
 \eeq
 where
 \beq
\vec g(\vec p_1,\vec p_2,\lambda) = 
i(p_2\epsilon_1\,\vec\nabla_{\vec p_2} -
p_1\epsilon_2\,\vec\nabla_{\vec p_1})\,
I(\vec q,\vec p_1,\vec p_2,\lambda)
\Bigr|_{\vec q=\vec p_1-\vec p_2}
\label{350}
 \eeq

Further calculations are moved to Appendix~A where they are performed
for a photon of mass $m_\gamma$. In the final Eq.~(\ref{a250}) the second term 
in the brackets vanishes since $F(1~+~iZ\alpha,1-iZ\alpha;2;\tilde x_0)$
diverges logarithmically, but $1-\tilde x_0 \propto m_\gamma^2 \to 0$.
The first term at $m_\gamma \to 0$ gets,
 \beq
F(iZ\alpha,-iZ\alpha;1;\tilde x_0)
_{m_\gamma\to 0} \longrightarrow 
\frac{1}{\left|\Gamma(1-iZ\alpha)\right|^2} =
{1\over N}\ .
\label{355}
 \eeq
 Then Eq.~(\ref{a250}) leads in the limit $m_\gamma\to 0$ to
 \beq
\vec d = \frac{4\pi}{Q^2}\,
(\epsilon_1\,\vec p_2 - \epsilon_2\,\vec p_1)\,
\left[\frac{Q^2}{(p_1+p_2+\nu)^2}\right]^{iZ\alpha}\ .
\label{360}
 \eeq
 Thus, we have proven that the amplitude is different from the Born one 
only by the phase factor.

\section{Coherent electroproduction of vector mesons}\label{coherent}

\subsection{VDM, a point-like nucleus}\label{point-like}

Apparently, the hadronic current depends on $Q^2$.
For instance, the VDM suggests
 \beq
\vec J(Q) = \vec J(0)\,\frac{m_V^2}{Q^2+m_V^2}\ ,
\label{400}
 \eeq
 where $m_V$ is the vector meson mass.
Including the propagator of the 
virtual photon  $1/Q^2$ it can be represented as,
 \beq
\frac{m_V^2}{Q^2\,(Q^2+m_V^2)} = \frac{1}{Q^2} - 
\frac{1}{Q^2+m_V^2}\ .
\label{410}
 \eeq Correspondingly, the new the vector $\vec d$ equals to the
difference between Eq.~(\ref{340}) and the same expression, but with
replacement $\nu \Rightarrow \sqrt{\nu^2-m_V^2}$. As a result, on top of
the phase factor the amplitude acquires another complex form factor,
 \beq
S(Q^2) = 1-\frac{\pi Z\alpha}{\sinh(\pi Z\alpha)}\,
x^{1-iZ\alpha}\,W(\kappa)\ ,
\label{420}
 \eeq
 where
 \beqn
W(\kappa) &=& F(iZ\alpha, -iZ\alpha;1;\kappa) -
iZ\alpha(1-\kappa)\,F(1+iZ\alpha,1-iZ\alpha;2;\kappa)\ ;
\label{430}\\
\kappa &=& \frac{Q^2}{Q^2+m_V^2}\ .
\nonumber
 \eeqn

\begin{figure}[tbh] 
\includegraphics{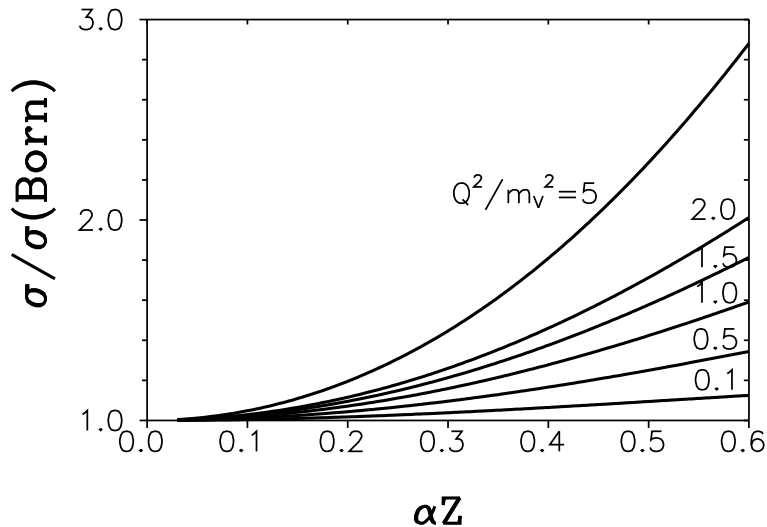} 
\begin{center} 
\vspace{7cm}
\parbox{13cm}
 {\caption[Delta]
 {\sl Deviation of the cross section from the Born approximation as 
function of $Z\alpha$ for different fixed values of $Q^2/m_V^2$.}
\label{vdm}} 
\end{center}
 \end{figure}

We calculated the ratio of the cross sections of reaction $l\,A\to
l'\,V\,A$ calculated with the distorted Coulomb wave functions and
in the Born approximation. The results are plotted it in Fig.~\ref{vdm} as
function of $Z\alpha$ for different values of $Q^2/m_V^2$.  
 Deviation from unity increases with $Q^2$, and is of course larger for
heavier nuclei.
Note that in the limit $Z\alpha\to 0$ the form factor Eq.~(\ref{420}) 
recovers the conventional VDM form Eq.~(\ref{400}) and the deviation from
the Born approximation vanishes, as Fig.~\ref{vdm} confirms.

If  the hadronic current contains a higher power of $Q^2$,
 \beq
J(Q) = J(0)\,\left(\frac{m_V^2}{m_V^2+Q^2}\right)^n\ ,
\label {440}
 \eeq
 it can be treated in a similar way provided that $n$ is integer. In this
case the previously obtained expressions should be differentiated in the
parameter $m_V^2$.

\subsection{More realistic hadronic current}\label{slope}

We assumed above for the sake of possibility of analytical calculations
that the hadron current $\vec J(Q)$ is independent of other variables,
but $Q$. This means, in particular, that the amplitude of virtual
coherent photoproduction $\gamma^*\,a \to V\,A$ is isotropic, i.e.
independent of the momentum transfer $\vec\Delta = \vec Q - \vec P_V$.
This is quite unrealistic suggestion, and a better form of the hadronic
current would be
 \beq
\vec J(Q,\vec\Delta) = 
\frac{\vec e_V\,m_V^2}{m_V^2+Q^2}\ 
\frac{1}{1+B\,\vec\Delta^2/2}\ ,
\label{500}
 \eeq where $\vec e_V$ is the polarization vector of the vector meson,
and $B$ is the slope of the transverse momentum distribution for the
reaction $\gamma^*A\to VA$.

In this case all multiple integrations in the amplitude of 
electroproduction of vector meson can be done analytically down to
the one dimensional integral,
 \beq
M(lA\to l^\prime VA) = \int\limits_0^1 dx\,
\vec e_V\cdot \vec h(\vec p_1,\vec p_2,\vec p_V, x)\ ,
\label{520}
 \eeq
 where $\vec h = \vec h_1 - \vec h_2$, and
 \beqn
\vec h_{1,2}(\vec p_1,\vec p_2,\vec p_V, x) &=& 
\frac{1}{2\,\omega_{1,2}}\,
\Biggl\{\left[ \epsilon_2\,p_1\,\vec\nabla_{p_1} -
\epsilon_1\,p_2\,\vec\nabla_{p_2} +
(\epsilon_1\vec p_2 - \epsilon_2\vec p_1)\,
\frac{\partial}{\partial\,\omega_{1,2}}\right]
\nonumber\\
&\times& J(\vec p_1,\vec p_2,\vec q,\omega_{1,2})
\Biggr\}_{\vec q=\vec p_1-\vec p_2-x\vec P_V}\ .
\label{530}
 \eeqn
 Here 
 \beqn
x &=& 1-\frac{uv}{wz}\ ;\nonumber\\
w &=& \vec q^{\,2} -\omega^2\ ;\nonumber\\
u &=& (\vec p_2+\vec q)^2 - 
(p_2+\omega)^2\ ;\nonumber\\
v &=& (p_1+\omega)^2 +
(\vec p_1-\vec q)^2\ ;\nonumber\\
z &=& (p_1+p_2+\omega)^2 -
(\vec p_1-\vec p_2-\vec q)^2\ ;\nonumber\\
\omega_1 &=& (1-x)\,\nu^2 -
x(1-x)\,p_V^2-x\,B/2\ ;\nonumber\\
\omega_2 &=& (1-x)(\nu^2-m_V^2) -
x(1-x)\,p_V^2-x\,B/2\ ;
\label{540}
 \eeqn
 and
\beqn
 J(\vec p_1,\vec p_2,\vec q,\omega_{1,2}) &=&
\int \frac{d^3r}{r}\,e^{i\omega r +i\vec q\cdot\vec r}\,
F(iZ\alpha,1,ip_1r-i\vec p_1\cdot\vec r)\,
F(iZ\alpha,1,ip_2r+i\vec p_2\cdot\vec r)
\nonumber\\ &=&
\frac{4\pi N}{w}\,\left({w\over z}\right)^{iZ\alpha}\,
F(iZ\alpha,1-iZ\alpha;1;x)\ .
\label{550}
 \eeqn

\begin{figure}[tbh] 
\includegraphics{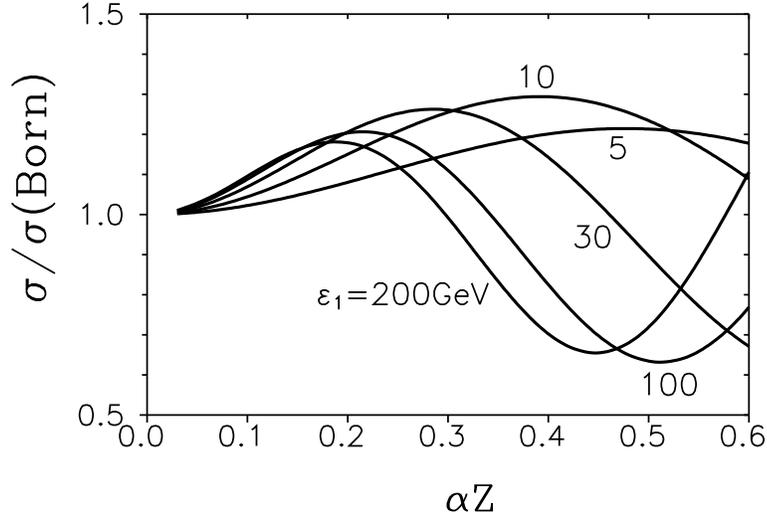} 
\begin{center} 
\vspace{7cm}
\parbox{13cm}
 {\caption[Delta]
 {\sl The same as in Fig.~\ref{vdm} but for different incident energies
$\epsilon_1=200,\ 100,\ 30,\ 10,$ and $5\ GeV$ at fixed $y=0.6$ and
$Q^2=m_\rho^2$.}
\label{vdm1}} 
\end{center}
 \end{figure}

We performed numerical calculations for forward production ($\vec\Delta =
0$) of transversely polarized vector mesons.  The ratio of the calculated
and Born cross sections $\sigma/\sigma_{Born} = |M|^2/|M_{Born}|^2$ is
depicted in Fig.~\ref{vdm1} for few values of $\epsilon_1$ at fixed
$y=(\epsilon_1-\epsilon_2)/\epsilon_1 = 0.6$ and $Q^2=m_\rho^2$.

The magnitude of the effect is probably overestimated since the nucleus
size is taken into account only in the production amplitude $\gamma^*A\to
VA$, while the wave functions of the incident and scattered leptons still
correspond to the potential of a point-like center (see next
Sect.~\ref{large-nucl}).  Nevertheless, these calculations reveal a new
effect, the Coulomb corrections can be either positive, like in the
previous case, or negative (compare with the results of calculations in
\cite{co}).

We have also studied how these Coulomb corrections vary as function of
$y$ and $Q^2$. While the magnitude of deviation from the Born cross
section is nearly independent of $y$, it substantially increases with
$Q^2$, keeping about the same shape of the dependence on $Z\alpha$.

\subsection{Effects of nuclear size on the lepton wave
functions}\label{large-nucl}

One should expect the Coulomb effects to be diminished when the size of
the nucleus is taken into account. Indeed, the V-meson interaction radius
is short and the beam lepton has to have about the same impact parameter 
as the
virtual photon or V-meson (the difference is $\sim 1/Q$).  Propagating
through the nucleus the lepton experiences a weaker electric field
compared to the case of a point-like Coulomb center. Correspondingly, the
Coulomb wave functions of the leptons must be corrected for the finite
size of the nucleus.

The general case of a massless lepton in an isotropic potential $V(r)$
has been solved and the radial wave functions, correct in all orders in
$Z\alpha$, have been found in \cite{lenz} within the quasiclassical WKB
approximation. They have been used later in \cite{knoll,ital,boffi} to
sum up the partial waves and build a formal expression for the lepton
wave functions,
 \beq
\Psi^\pm(\vec p,\vec r) = e^{\pm i\delta_{1/2}}\,
\eta(r)\,e^{\pm ib(\vec J^{\,2}-3/4)}\,
e^{i\vec p\cdot\vec r\,\eta(r)}\,u(p)\ ,
\label{600}
 \eeq
where $\vec J = \vec L + \vec\sigma/2$ is the operator of the total angular
momentum;
 \beqn
b &=& -\frac{1}{2\,p^2}
\int\limits_0^\infty dr\,{1\over r}\,
\frac{dV(r)}{dr}\ ;\nonumber\\
\delta_{1/2} &=& - Z\alpha\,\ln(p\,R_A) -
\int\limits_0^{R_A}dr\,V(r) + b\ ;\nonumber\\
\eta(r) &=& \frac{1}{\rho\,r}
\int\limits_0^r dr^\prime\, \Bigl[p-V(r)\Bigr]\ ;
\label{610}
 \eeqn
and $u(p)$ is the 4-spinor of the lepton. Eq.~(\ref{600}) is a 
generalization of the Furry's wave functions Eq.~(\ref{110}).

Apparently, such an expression with the operator $\vec J$ in the exponent
is not easy to use for practical applications. This is why only the first
two terms of the order of $(Z\alpha)^0$ and $(Z\alpha)^1$ in the
expansion of the exponential in (\ref{600})  have been considered in
\cite{knoll,ital,boffi}. We skip here those lengthy expressions, but
apply the procedure developed in \cite{knoll,ital} to our case. We assume
a homogeneous charge distribution inside a sphere of nuclear radius $R_A$
and use the approximation $\epsilon_{1,2}R_A \gg 1$ and $\theta_{12} \ll
1$, where $\theta_{12}$ is the lepton scattering angle. Then the ratio of
the amplitude modified by the Coulomb field to the Born one takes the
form,
 \beq
\frac{M(lA\to l^\prime VA)}
{M_{Born}(lA\to l^\prime VA)} = 
1 + i\,\frac{3B\,Z\alpha}{R_A^2} = 
1 + i\,\frac{3\,Z\alpha}{5}\ .
\label{620}
 \eeq
 Here $B$ is the slope parameter of the differential cross section
introduced in (\ref{500}). It is related to the mean charge nuclear
radius squared,
 \beq
B = {1\over3}\,\la r_{ch}^2\ra_A =
{1\over5}\,R_A^2\ .
\label{630}
 \eeq

Thus, deviation from the Born cross section,
 \beq
\frac{\sigma(lA\to l^\prime VA)}
{\sigma_{Born}(lA\to l^\prime VA)} =
1 + \frac{9}{25}\,(Z\alpha)^2\ ,
\label{640}
 \eeq This is a sizeable correction for heavy nuclei, for example, the
modified cross section on lead is $13\%$ higher than the Born one.

Surprisingly, the correction Eq.~(\ref{640}) does not expose any
dependence on reaction kinematics, on the contrary to the results of
previous sections. It is probably a consequence of the higher order terms
in $Z\alpha$ missed in this calculations. How to elaborate with those
terms is still a challenge which we leave for further studies. The
purpose of the present estimates is to see whether the finiteness of
nuclear size can substantially diminish the effect of long-range Coulomb
forces. Apparently not, the deviation is still sizeable. One should
calculate at least the next term $O(Z^2\alpha^2)$ to make it sure that
the correction Eq.~(\ref{620}) is not just a phase. We believe that such
a possibility is very improbable, it happened so far only in the case of
elastic scattering, and in the special artificial case when the hadronic
current is a constant (Sect.~\ref{theorem}).  In both cases the Furry
wave functions for a point-like Coulomb center are used. The example of
Sect.~\ref{point-like} demonstrates that the even a simplest $Q^2$
dependence of the hadronic current leads to an amplitude which consists
of few terms having different phases, and the cross section changes.

\section{Eikonal approximation}\label{eikonal}

A different approach to the problem of distortion of the lepton wave
functions in the Coulomb field of an extended nucleus is the eikonal
approximation of Bjorken, Kogut and Soper (BKS) \cite{bjorken}.  It is
not clear how precise this approximation is, the best way to figure it
out is to compare its results with the known exact solution in the case
when it is available. The exact cross section has been calculated by
Davies, Bethe and Maximon (DBM) \cite{dbm} for photoproduction of
$e^+e^-$ pairs and Bethe and Maximon for bremsstrahlung \cite{maximon}.
Using the BKS approach we calculate in \ref{b} the cross section for this
reaction and arrive at exactly the same Coulomb correction to the Born
cross section as it has been found in \cite{dbm,maximon}. This success
can serve as an argument that the eikonal approximation is rather 
accurate.

In the eikonal approximation the wave functions of spinless leptons read,
 \beqn
\Psi^+(\vec p_1,\vec r) &=& 
\exp\Bigl[i\,\vec p_1\cdot\vec r - 
i\,\chi_1(\vec r)\Bigr]\ ;\nonumber\\
\Psi^-(\vec p_2,\vec r) &=&                                                                    
\exp\Bigl[i\,\vec p_2\cdot\vec r +
i\,\chi_2(\vec r)\Bigr]\ ,                                     
\label{700}
 \eeqn
 Here the phase shifts  are
 \beqn
\chi_1(\vec r) &=& 
\int\limits_{-\infty}^{z} dz'\,V(\vec b,z')\ ,
\nonumber\\
\chi_2(\vec r) &=&                         
\int\limits_z^{\infty} dz'\,V(\vec b,z')\ .
\label{710}
 \eeqn
 At high energies, $\epsilon_{1,2} \gg \sqrt{Q^2}$, the vectors
$\vec p_1$, $\vec p_2$ and $\vec p_1- \vec p_2$ are nearly
parallel. We chose the axis $z$ along $\vec p_1$, correspondingly the
vector $\vec r = (\vec b,z)$ has projection $\vec b$ to the normal plane.

For the long-range Coulomb field, $V(r)_{r\to\infty} = \pm\,Z\alpha/r$ ,
the integrals in (\ref{710}) are strictly speaking divergent. To fix the
problem we introduce an infra-red cut-off,
 \beq
V(r)\Bigr|_{r\to\infty} = \pm\,\frac{Z\alpha}{r}
\ e^{-\lambda r}\,\left(1-e^{-\mu r}\right)\ ,  
\label{720}
 \eeq 
 where the last factor corresponds to the pole form of the nuclear 
formfactor with
 \beq
\mu^{-2} = \frac{\la r_{ch}^2\ra_A}{6}\ .
\label{725} 
 \eeq
 
One may interpret $\lambda$ in (\ref{720}) as an effective photon mass,
or (better justified)  as the inverse screening radius of the nuclear
Coulomb field by the atomic electrons. As soon as $1/\lambda \gg R_A$,
any variation of the value of $\lambda$ may lead only to $r$-independent
additive corrections to the phase shifts $\chi_{1,2}(\vec r)$ what does
not affect the value of the cross section we are interested in.

Utilizing the same realistic form of the hadronic current for
electroproduction of vector mesons as in Sect.~\ref{slope} we get the
following expression for the amplitude,
 \beq
M(lA\to l^\prime VA) = 
\int\limits_0^1 dx\ \vec e_V\cdot
\vec f(\vec p_1,\vec p_2,\vec p_V,x)\ ,
\label{730}
 \eeq
where
 \beqn
\vec f = \vec f_1 - \vec f_2\ ,
\nonumber
 \eeqn
and
 \beqn
\vec f_{1,2}(\vec p_1,\vec p_2,\vec p_V,x) &=&
\frac{1}{2\omega_{1,2}}\,
\frac{\partial}{\partial\omega_{1.2}}
\int\frac{d^3r}{r}\,
\left\{\Bigl[\epsilon_1\vec p_2 -\epsilon_2\vec p_1\Bigr] -
\Bigl[\epsilon_1\vec\nabla\,\chi_2(\vec r) +
\epsilon_2\vec\nabla\,\chi_1(\vec r)\Bigr]\right\}
\nonumber\\ &\times&
\exp\Bigl[i\vec\kappa\vec r - 
i\chi_1(\vec r) - i\chi_2(\vec r) +
i\,\omega_{1,2}\,r\Bigr]\ ,
\label{740}
 \eeqn 
  where $\vec\kappa = \vec p_1-\vec p_2 -x\,\vec p_V$.

Note that the $x$-dependence of this expression comes via $\omega_{1,2}$
defined in (\ref{540}). The derivatives $\vec\nabla\chi_{1,2}(r)$ in
(\ref{740}) are the momenta transferred by the Coulomb field to the
initial and final leptons. In the case of an extended nucleus they are of
the order of $Z\alpha/R_A$.

Using relation
 \beq
\Bigl|\epsilon_1\vec p_2 -\epsilon_2\vec p_1\Bigr|=
\sqrt{\epsilon_1\epsilon_2\,Q^2}\ ,
\label{750}
 \eeq
 we see that for $Q^2\gg R_A^{-2}$ the second term in curly brackets in
(\ref{740}) is much smaller than the first one and can be neglected.
Then Eq.~(\ref{710}) takes the form,
 \beq
\chi_1(\vec r) + \chi_2(\vec r) =
\int\limits_{-\infty}^{\infty}dz\,V(\vec b,z) =
\chi(\vec b) = \pm 2Z\alpha\Bigl\{K_0(\lambda b) -
K_0[(\mu+\lambda) b]\Bigr\}\ ,
\label{760}
 \eeq
 where plus and minus correspond to different signs of the potential
Eq.~(\ref{720}).
Since the sum $\chi_1(\vec r) + \chi_2(\vec r)$ is independent of $z$
one can explicitly perform integration over $z$ in Eq.~(\ref{740}) using the
following relation,
 \beq
\int\limits_{-\infty}^{\infty} \frac{dz}{r}\,
\exp(i\,\kappa_L\,z + i\,\omega\,r) = 
2\,K_0\Bigl(b\,\sqrt{\kappa_L^2-\omega^2}\Bigr)\ ,
\label{770}
 \eeq
 where $K_0(x)$ is the modified Bessel function, $\omega=\omega_{1,2}$,
and $\kappa_{L,T}$ are the longitudinal and transverse components of 
$\vec\kappa$.

\begin{figure}[tbh] 
\includegraphics{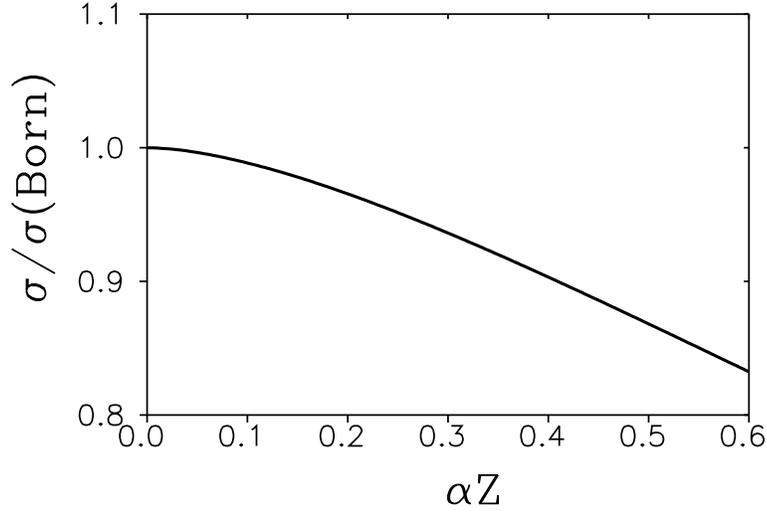} 
\begin{center} 
\vspace{7cm}
\parbox{13cm}
 {\caption[Delta]
 {\sl The same as in Fig.~\ref{vdm}, but calculated in the eikonal
approximation. The curve is calculated at $Q^2>1\,\GeV^2$ and is 
independent of $Q^2$ within the range of coherence $q_L\lsim1/R_A$.}
\label{eikonal-fig}} 
\end{center}
 \end{figure}

Thus, the problem of calculation of the amplitude (\ref{730}) is reduced to
integrals of the form,
\beq
\int\limits_0^1 dx \int\limits_0^{\infty}
db\,b\,K_0\Bigl(b\,\sqrt{\kappa_L^2-\omega^2}\Bigr)\,
e^{i\chi(b)}\,J_0(\kappa_T b)\ , 
\label{780}
 \eeq 
 which we calculate numerically. Our results for ratio of the calculated
and Born cross sections are depicted in Fig.~\ref{eikonal-fig} as
function of $Z\alpha$. 
 It turns out that the cross section is smaller than the Born one, and
the difference is $Q^2$ independent at $Q^2>1\,\GeV^2$. However, the
range of $Q^2$ correlates with the photon energy $\nu$ to satisfy the
condition of coherence $q_L=(Q^2 + m_V^2)/2\nu \lsim 1/R_A$.

\section{Conclusions and outlook}\label{conclusions}

The Born approximation employed in all analyses of experimental data has
no justification for heavy nuclei with $Z\alpha\sim 1$. Therefore, the
accuracy of experimental results for nuclear effects in DIS may be
essentially affected by this theoretical uncertainty.

Radiative corrections are known to provide a substantial contribution to
the DIS cross section. Usually they can be calculated rather accurately
\cite{bardin}, except one related to the modification of the lepton wave
function by long-range Coulomb forces. This correction violating the Born
approximation is usually ignored, and for a good reason:  it is a very
difficult task to calculate it. Nevertheless this problem should be
challenged rather than 'swept under the carpet'.

A certain progress is made in the present paper towards the calculation
of the DIS cross section using the lepton wave functions modified by the
long-range Coulomb forces. We start with the simplest assumption that the
hadronic current is independent of $Q^2$ and prove that the DIS amplitude
acquires only a phase. In this case the Coulomb corrections do not modify
the Born cross section.

This conclusions, however, changes after a $Q^2$ dependence is introduced
into the hadronic current. We demonstrate that on the example of coherent
leptoproduction of vector mesons off a point-like nucleus which we treat
within VDM. Of course VDM should not be used at high $Q^2$ where color
transparency is important. However, we just want to estimate the scale of
the correction, while precise calculations ready to use in an analysis of
data is still a challenge.

Further steps beyond the approximation of a point-like target lead to a
conclusion that in the case of an extended nucleus the Coulomb
corrections are still important and can affect the existing experimental
results for nuclear effects in DIS which are based on the Born
approximation. To do more precise calculations one needs to know the
lepton wave functions in the Coulomb field of an extended nucleus which
are currently available only in the first order in $Z\alpha$.

We also use the eikonal approximation which is better designed for the
case of extended nuclei. First we demonstrate (\ref{b}) that this
approximation well reproduces the known exact results when they are
available. Then we apply this method to the process of coherent
electroproduction of vector mesons and again arrive at a sizeable
correction to the Born cross section.

We conclude that the long-range Coulomb forces may significantly modify
the DIS cross section on heavy nuclei compared to the widely used Born
approximation. Further progress in this direction should help to make
experimental results for nuclear effects in DIS more reliable.

Although originally our work has been motivated by the unusual nuclear
effects in DIS observed in the HERMES experiment \cite{hermes} (see
Introduction), we did not find anything special about the energy and $x$
range of this data.  Whatever deviations from the Born approximation
happen due to the long-range Coulomb forces, they should have similar
magnitude either at the energy of HERMES, or NMC experiments.

\medskip \noindent {\bf Acknowledgment}: We are grateful to Andreas
Sch\"afer who initiated this work for interesting and valuable
discussions. This work has been partially supported by a grant from the
Gesellschaft f\"ur Schwerionenforschung Darmstadt (GSI), grant
No.~GSI-OR-SCH, and by the European Network: Hadronic Physics with
Electromagnetic Probes, Contract No.~FMRX-CT96-0008.

 \def\appendix{\par
 \setcounter{section}{0} \setcounter{subsection}{0}
 \def\thesection{Appendix \Alph{section}}
\def\thesubsection{\Alph{section}.\arabic{subsection}}
\def\theequation{\Alph{section}.\arabic{equation}}
\setcounter{equation}{0}}                                                      

 \appendix 

\section{Calculation of the vector \boldmath$\vec d$}\label{a}

We present here the details of calculations of the vector $\vec d$ in
Eq.~(\ref{340}). Since we are going to use the results also in
Sect.~\ref{point-like} where the amplitude contains two terms in
Eq.~(\ref{410}), we will treat the photon as a massive particle with mass
$m_\gamma$ which is either zero [the first term in (\ref{410})], or
equals to $m_V$ [the first term in (\ref{410})]. Correspondingly, in all
equations of Sect.~\ref{theorem} one should replace $\nu \Rightarrow
\tilde\nu=\sqrt{\nu^2-m_\gamma^2}$.

The expression for the vector $g(\vec p_1,\vec p_2,\lambda)$ in
Eq.~(\ref{350}) can be represented as
 \beqn
&& \vec g(\vec p_1,\vec p_2,\lambda) = 
\left[\left(p_2\epsilon_1\,\vec\nabla\cdot\vec p_2 -
p_1\epsilon_2\,\vec\nabla\cdot\vec p_1\right)\,
I(\vec q,\vec p_1,\vec p_2,\lambda)\right]_{\vec q=
\vec p_1-\vec p_2}\nonumber\\
&+& \frac{4\,\pi\,i\,N}
{Q^2+m_\gamma^2+\lambda^2-2i\tilde\nu}\ 
\left[\frac{Q^2+m_\gamma^2+\lambda^22-2i\tilde\nu}
{(p_1+p_2+\tilde\nu+i\lambda)^2}\right]^{iZ\alpha}
\nonumber\\ &\times&
\left\{-\,\frac{2iZ\alpha(\epsilon_1\vec p_2 -
\epsilon_2\vec p_1)}{p_1+p_2+\tilde\nu+i\lambda}\ 
\Phi(\tilde x) - 2\,\left(\frac{\epsilon_1}{D_1} -
\frac{\epsilon_2}{D_2}\right)\,
(p_1\vec p_2 - p_2\vec p_1)(1-\tilde x)\,
\Phi^\prime(\tilde x)\right\}\ .
\label{a10}
 \eeqn
 The following notations are used here,
 \beqn
\Phi(\tilde x) &=& F(iZ\alpha,1-iZ\alpha;1;\tilde x)\ ;
\label{a20}\\
\Phi^\prime(\tilde x) &=& \frac{d}{d\,\tilde x}\,
\Phi(\tilde x)\ ;
\label{a30}\\
\tilde x &=& x_{\vec q=\vec p_1-\vec p_2} =
\frac{Q^2-\Delta^2}{Q^2+m_\gamma^2+\lambda^2-2i\tilde\nu\lambda}\ ;
\label{a40}\\
Q^2 &=& \vec q^2 - \nu^2\ ;\nonumber\\
\Delta^2 &=& (p_1-p_2)^2-\nu^2\ ;
\label{a50}\\
D_1 &=& u_{\vec q=\vec p_1-\vec p_2} =
p_1^2 - (p_2+\tilde\nu+i\lambda)^2\ ;\nonumber\\
D_2 &=& v_{\vec q=\vec p_1-\vec p_2} =
(p_1+\tilde\nu+i\lambda)^2-p_2^2\ .
\label{a60}
 \eeqn

Disregarding the difference between $p_{1,2}$ and $\epsilon_{1,2}$
(equivalent to the dropping off the corrections of the order of
$m_\gamma^2/p_{1,2}^2$) we can write,
 \beq
\frac{2\epsilon_1}{D_1} - \frac{2\epsilon_2}{D_2} =
\frac{2(\tilde\nu+i\lambda)}
{(p_1-p_2)^2+(\lambda-i\tilde\nu)^2} -
\frac{2}{p_1+p_2+\nu+i\lambda}\ .
\label{a70}
 \eeq
Making use of this relation we get,
 \beqn
\vec g(\lambda) &=& \frac{4\pi N\,(\epsilon_1\vec p_1 -
\epsilon_2\vec p_1)}
{Q^2+m_\gamma^2+\lambda^2-2i\tilde\nu\lambda}\ 
\left[\frac{Q^2+m_\gamma^2+\lambda^2-2i\tilde\nu\lambda}
{(p_1+p_2+\tilde\nu+i\lambda)2}\right]^{iZ\alpha}
\nonumber\\ &\times&
\left[\frac{2(\lambda-i\tilde\nu)}
{p_1-p_2)^2 + (\lambda-i\tilde\nu)^2}\ 
(1-\tilde x)\,\Phi^\prime(\tilde x) + 
\frac{2Z\alpha\,\Phi(\tilde x)+ 2i(1-\tilde x)\,
\Phi^\prime(\tilde x)}
{p_1+p_2+\tilde\nu+i\lambda}\right]\ .
\label{a80}
 \eeqn
 Notice that
 \beq
\frac{2(\lambda-i\tilde\nu)(1-\tilde x)}
{(\vec p_1-\vec p_2)^2 + (\lambda+i\tilde\nu)^2} = 
\frac{2(\lambda-i\nu)}
{Q^2+m_\gamma^2+\lambda^2-2i\nu\lambda} = 
\,-\,\frac{1}{\tilde x}\,\frac{\partial\tilde x}
{\partial\lambda}\ ,
\label{a90}
 \eeq
and
 \beq
2Z\alpha\,\Phi(\tilde x) + 2i(1-\tilde x)\,
\Phi^\prime(\tilde x) = i(Z\alpha)^2\,
\tilde\Phi(\tilde x)\ ,
\label{a100}
 \eeq
 where
 \beq
\tilde\Phi(\tilde x) = F(1+iZ\alpha,1-iZ\alpha;2;\tilde x)\ .
\label{a110}
 \eeq
 Therefore, the vector $\vec g(\lambda)$ gets the form,
 \beqn
\vec g(\lambda) &=& \frac{4\pi N(\epsilon_1\vec p_2-
\epsilon_2\vec p_1)}{Q^2}\ 
\left[\frac{Q^2+m_\gamma^2+\lambda^2-2i\tilde\nu\lambda}
{p_1+p_2+\tilde\nu+i\lambda)^2}\right]^{iZ\alpha}
\nonumber\\ &\times&
\left[ -\,\frac{\partial\Phi}{\partial\lambda} +
\frac{i(Z\alpha)^2\,\tilde x\,\tilde\Phi(\tilde x)}
{p_1+p_2+i\lambda+\tilde\nu}\right]\ .
\label{a120}
 \eeqn 
 Here the second term in the brackets of the last factor is small, $\sim
\theta^2\ln(1/\theta)$, relative to the first term (see below),
therefore, it can be neglected for the small scattering angles $\theta\ll
1$. In this case the problem under consideration is reduced to the
integral,
 \beq
\int\limits_0^\infty d\lambda\,\vec g(\lambda) =
\frac{4\pi N}{Q^2}\,
(\epsilon_1\vec p_2-\epsilon_2\vec p_1)\,L
\label{a125}
 \eeq
 where
 \beq
L = \int\limits_0^\infty d\lambda\,t^{iZ\alpha}\,
\frac{\partial\Phi}{\partial\lambda}\ ,
\label{a130}
 \eeq
and 
 \beq
t = \frac{Q^2+m_\gamma^2+\lambda^2 - 2i\nu\lambda}
{p_1+p_2+\tilde\nu+i\lambda}\ .
\label{a140}
 \eeq
Integrating in Eq.~(\ref{a130}) by parts we get,
 \beq
L = t^{iZ\alpha}\,\Phi(\tilde x)
\Bigr|^{\lambda=\infty}_{\lambda=0} -
\int\limits_0^\infty d\lambda\,\frac{\partial\,t^{iZ\alpha}}
{\partial\,\lambda}\, \Phi(\tilde x)\ .
\label{a150}
 \eeq
Since $\tilde x=0$ at $\lambda\to\infty$ and $\Phi(0)=1$,
we have,
 \beq
t^{iZ\alpha}\,\Phi(\tilde x)\Bigr|_{\lambda=0}
^{\lambda=\infty} = (-1)^{iZ\alpha} -
t_0^{iZ\alpha}\,\Phi(\tilde x_0)\ ,
\label{a160}
 \eeq
 where
 \beqn
t_0 &=& t(\lambda=0)=\frac{Q^2+m_\gamma^2}
{(p_1+p_2+\tilde\nu)^2}\ ;\nonumber\\
\tilde x_0 &=& \tilde x(\lambda=0) =
\frac{Q^2}{Q^2+m_\gamma^2}\ .
\label{a170}
 \eeqn

To calculate the rest integrals we notice that the value of $\tilde x$
is tiny, except the region of $\lambda < (Q^2+m_\gamma^2)/(2\tilde\nu)
\ll 1$. Then, we split the integral in Eq.~(\ref{a150}) into two parts,
 \beq
\int\limits_0^\infty d\lambda \ =\ 
\int\limits_0^{\lambda_1} d\lambda\ \ +\ \ 
\int\limits_{\lambda_1}^\infty d\lambda\ ,
\label{a180}
 \eeq
 where we chose $(Q^2+m_\gamma^2)/(2\tilde\nu)\ll \lambda_1\ll \nu$, so
that $|\tilde x|\ll 1$ in the second term in (\ref{a180}). Therefore, we
can fix $\Phi(\tilde x)=1$ in the second integral in (\ref{a180}) and
get,
 \beq
\int\limits_{\lambda_1}^\infty d\lambda\,
\frac{\partial\,t^{iZ\alpha}}{\partial\,\lambda}\,
\Phi(\tilde x) \approx
t^{iZ\alpha}\Bigr|_{\lambda_1}^\infty +
O\left(\frac{\lambda_1}{\nu}\right) =
(-1)^{iZ\alpha} - t_1^{iZ\alpha} + 
O\left(\frac{\lambda_1}{\nu}\right)\ ,
\label{a190}
 \eeq
 where $t_1=t(\lambda_1)$.

In the first integral in (\ref{a180}) we can make 
use of smallness of $\lambda$ in the expression (\ref{a140})
and rewrite it as,
 \beq
t = \frac{Q^2+m_\gamma^2-2i\lambda\tilde\nu}
{(p_1+p_2+\tilde\nu)^2} =
\frac{\tilde x_0\,t_0}{\tilde x}\,{1\over x}\ .
\label{a200}
 \eeq
Then the first integral in (\ref{a180}) takes the form,
 \beqn
\int\limits_0^{\lambda_1} d\lambda\,
\frac{\partial\,t^{iZ\alpha}}{\partial\,\lambda}\,
\Phi(\tilde x) &=& -\,iZ\alpha\,
(\tilde x_0\,t_0)^{iZ\alpha}\,
\int\limits_{\tilde x_0}^{\tilde x_1} d\tilde x\,
\tilde x^{-1-iZ\alpha}\,\Phi(\tilde x)\nonumber\\
&=&
(\tilde x_0\,t_0)^{iZ\alpha}\,
\left[\tilde x^{-iZ\alpha}\,
F(iZ\alpha,iZ\alpha;1;\tilde x)\right]
\Biggr|_{\tilde x_0}^{\tilde x_1}\nonumber\\
&=& t_1^{iZ\alpha} - t_0^{iZ\alpha}\,
F(iZ\alpha,iZ\alpha;1;\tilde x_0) +
O\left(\frac{\lambda_1}{\nu}\right)\ .
\label{a210}
 \eeqn
Here we used the definition of $\Phi(\tilde x)$ from Eq.~(\ref{a20}).

Now, taking into account that $|\tilde x(\lambda=\lambda_1)|\ll 1$
we eventually arrive at the following expression for the integral 
Eq.~(\ref{a130}),
 \beq
L = t_0^{iZ\alpha}\,\left[F(iZ\alpha,iZ\alpha;1;\tilde x_0) -
F(iZ\alpha,1-iZ\alpha;1;\tilde x_0)\right]
\ .
\label{a220}
 \eeq

Taking into account the relation,
 \beq
F(iZ\alpha,iZ\alpha;1;\tilde x_0)=
F(iZ\alpha,-iZ\alpha;1;\tilde x_0) +
iZ\alpha\,\tilde x_0\,F(1+iZ\alpha,1-iZ\alpha;2;\tilde x_0)\ ,
\label{a230}
 \eeq
we can integrate $\vec g(\lambda)$ as,
 \beq
\int\limits_0^\infty d\lambda\, g(\lambda) =
\frac{4i\pi(\epsilon_1\vec p_2-\epsilon_2\vec p_1)}
{Q^2+m_\gamma^2}\ 
\left[\frac{Q^2+m_\gamma^2}
{(p_1+p_2+\tilde\nu)^2}\right]^{iZ\alpha}\,
iZ\alpha\,F(1+iZ\alpha,1-iZ\alpha;2;\tilde x_0)\ .
\label{a240}
 \eeq 
 Substituting this expression into Eq.~(\ref{340}) we eventually arrive
at the final expression for $\vec d$,
 \beqn
\vec d &=& \frac{4\pi\,N}{Q^2+m_\gamma^2}\,
(\epsilon_1\vec p_2-\epsilon_2\vec p_1)\,
\left[\frac{Q^2+m_\gamma^2}
{(p_1+p_2+\tilde\nu)^2}\right]^{iZ\alpha}
\nonumber\\ &\times&
\left[F(iZ\alpha,-iZ\alpha;1;\tilde x_0) -
iZ\alpha(1-\tilde x_0)\,
F(1+iZ\alpha,1-iZ\alpha;2;\tilde x_0)\right]\ .
\label{a250}
 \eeqn

\setcounter{equation}{0}

\section{Eikonal approximation versus the exact calculations}\label{b}

Here we calculate the total cross section of photoproduction
of electron-positron pairs off atoms at high energies relying
upon the eikonal approximation of ref. \cite{bjorken}
and compare with the exact results of \cite{dbm}.

According to \cite{bjorken} the photoproduction cross section has 
the factorized form,
 \beq
\sigma(\gamma Z\to e^+e^-Z) =
\int\limits_0^1 du \int d^2\rho\,
\sigma(\rho, u)\,
\Bigl|\Psi(\vec \rho,u)\Bigr|^2\ ,
\label{b10}
 \eeq
where $\sigma(\rho, u)$ is the cross section of interaction of the $e^+e^-$
dipole with the atom which depends on the transverse separation 
$\vec\rho$ and the fraction $u=(E+p_L)/2\omega$ of the 
photon light-cone momentum carried by the electron. 
$\Psi(\vec \rho,u)$ is the $e^+e^-$ wave function of the photon,
 \beq
\Bigl|\Psi(\vec \rho,u)\Bigr|^2 = 
\frac{\alpha\,m^2}{2\pi^2}\,
\left\{K_0^2(m\rho) +
\Bigl[u^2+(1-u)^2\Bigr]\,
K_1^2(m\rho)\right\}\ ,
\label{b20}
 \eeq
where $m$ is the electron mass, $\alpha$ is the fine structure constant.

If the atom remains intact, then according to \cite{bjorken},
 \beq
\sigma(\rho,u) = 2\int d^2b\,
\left\{1-\exp\Bigl[i\,\chi(\vec b-u\vec\rho) -
i\,\chi(\vec b + \vec\rho - u\vec\rho)\Bigr]\right\}\ ,
\label{b30}
 \eeq
where $\chi(\vec b)$ is defined in (\ref{760}).

Replacing in (\ref{b30}) $\vec b +(1/2-u)\vec\rho \Rightarrow \vec b$
we arrive at,
 \beq
\sigma(\rho,u) = 2\int d^2b\,
\left\{1-\exp\Bigl[i\,\chi(\vec b_+) -                            
i\,\chi(\vec b_-)\Bigr]\right\}\ ,                         
\label{b40}             
 \eeq
 where $\vec b_\pm=\vec b \pm \vec\rho/2$. Thus, we conclude that the
dipole cross section Eq.~(\ref{b40}) depends only on the transverse
separation $\rho$.

The phase shifts $\chi(\vec b_\pm)$ can be expressed in terms of the
transverse density of electrons in the atom $n(s)$,
 \beq
\chi(\vec b_\pm) =
2Z\alpha \int\limits_{b_\pm}^{\infty} 
ds\,s\,n(s)\,\ln\left(\frac{s}{b_\pm}\right)\ ,
\label{b50}
 \eeq
where $b_\pm = |\vec b_\pm|$ and the density is normalized as,
 \beq
\int_0^\infty ds\,s\,n(s) = 1\ .
\label{b55}\
 \eeq

The atomic size $R_Z \sim 1/(m\alpha Z^{1/3})$ is much 
larger than the transverse $e^+e^-$ separation $\rho \sim 1/m$,
therefore, one can split the integral in (\ref{b40}) into two parts,
 \beq
\sigma(\rho) = 2\int\limits_0^\infty
db\,b\,\int\limits_0^{2\pi} d\phi\,
\Bigl[1-\exp(i\chi_+-i\chi_-)\Bigr] = 
\sigma_1(\rho) + \sigma_2(\rho)\ ,
\label{b60}
 \eeq
where $\phi$ is the azimuthal angle between $\vec b$ and $\vec\rho$; 
$\chi_\pm=\chi(\vec b_\pm)$, and 
 \beqn
\sigma_1(\rho) &=& 2\int\limits_0^{b_0}                                 
db\,b\,\int\limits_0^{2\pi} d\phi\,
\Bigl[1-\exp(i\chi_+-i\chi_-)\Bigr]
\label{b70}\\
\sigma_2(\rho) &=& 2\int\limits_{b_0}^\infty 
db\,b\,\int\limits_0^{2\pi} d\phi\,
\Bigl[1-\exp(i\chi_+-i\chi_-)\Bigr] .
\label{b80}                                   
 \eeqn
 We chose the value of $b_0$ satisfying the condition 
 \beq
{1\over m} \ll b_0 \ll R_z\ .
\label{b90}
 \eeq

Starting with Eq.~(\ref{b80}) for $\sigma_2(\rho)$ we note that the mean
value of the $e^+e^-$ separation is much smaller than the impact parameter,
$$ \rho \sim {1\over m} \ll b_0 \leq b\ .$$
Therefore,
 \beqn
\chi_+ - \chi_- &=& 
\chi\Bigl(\vec b +\frac{\vec\rho}{2}\Bigr) -
\chi\Bigl(\vec b -\frac{\vec\rho}{2}\Bigr) =
\vec\rho\cdot\vec\nabla_b\,\chi(\vec b)
+ O\Biggl(\frac{\rho^3}{b^3}\Biggr)\nonumber\\&\approx& 
\rho\,\cos \phi\ \frac{d\chi(b)}{d\,b} =
2Z\alpha\cos \phi\,\frac{\rho}{b}\,
\int\limits_b^\infty ds\,s\,n(s)\ .
\label{b100}
 \eeqn
Correspondingly,
 \beqn
\sigma_2(\rho) &=& 4\pi\int\limits_{b_0}^\infty
db\,b\,\left[1-J_0\Biggl(\rho\,
\frac{d\chi}{d\,b}\Biggr)\right]\nonumber\\
&=& 4\pi(Z\alpha)^2\rho^2 \int\limits_{b_0}^\infty
\frac{d\,b}{b}\,\left[\int\limits_{b}^\infty
ds\,s\,n(s)\right]^2 + O\Biggl(\frac{\rho^4}{R_z^4}\Biggr)\ .
\label{b110}
 \eeqn
Integrating this expression by parts we get,
 \beqn
\int\limits_{b_0}^\infty
\frac{d\,b}{b}\,\left[\int\limits_{b}^\infty      
ds\,s\,n(s)\right]^2 &=&
\ln\Biggl({1\over b_0}\Biggr)\,
\left[\int\limits_{b_0}^\infty      
ds\,s\,n(s)\right]^2\nonumber\\ &-&
2\int\limits_{b_0}^\infty db\,b\,
\ln\Biggl({1\over b\,}\Biggr)\,
n(b)\,\int\limits_{b}^\infty 
ds\,s\,n(s)\ .
\label{b120}
 \eeqn

The bottom limits $b_0$ of the integrals in the right-hand side (r.h.s.)  
of (\ref{b120}) can be replaced by zero, $b_0 \Rightarrow 0$, with an
accuracy of the order of $b_0^2/R_Z^2 \ll 1$. Then, taking into account
the normalization condition Eq.~(\ref{b55}) we get,
 \beq
\sigma_2(\rho) = 4\pi(Z\alpha)^2\,\rho^2\,
\left[\ln\Biggl(\frac{a}{b_0}\Biggr) +
O\Biggl(\frac{b_0^2}{R_Z^2}\Biggr)\right]\ ,
\label{b130}
 \eeq
 where
 \beq
\ln a = -2\int\limits_0^\infty db\,b\,
\ln\Biggl({1\over b\,}\Biggr)\,n(b)\int\limits_{b}^\infty       
ds\,s\,n(s)\ .
\label{b140}                                   
 \eeq                                       
Apparently, the value of $a\sim R_Z$ depends only on the details of 
the atomic structure contained in the electron density distribution $n(s)$.

Now we turn to the first term $\sigma_1(\rho)$ in (\ref{b60}).
Since $b\leq b_0\ll R_Z$, also $b_\pm=|\vec b \pm \vec\rho/2|
\leq b_0 \ll R_Z$. Making use of these relations we get,
 \beqn
\chi_\pm &=& 2Z\alpha
\int\limits_{b_\pm}^\infty ds\,s\,n(s)\,
\ln\Biggl(\frac{s}{b_\pm}\Biggr)\nonumber\\
&=& 2Z\alpha \int\limits_0^\infty
ds\,s\,n(s)\,\ln\Biggl(\frac{s}{b_\pm}\Biggr)
+ O\Biggl(\frac{b_\pm^2}{R_Z^2}\Biggr)\ .
\label{b150}
 \eeqn
Consequently,
 \beqn
\chi_+ - \chi_- &=& 2Z\alpha\,
\ln\Biggl(\frac{b_-}{b_+}\Biggr)
\int\limits_0^\infty ds\,s\,n(s) +
O\Biggl(\frac{b_\pm^2}{R_Z^2}\Biggr)\nonumber\\
&\approx& Z\alpha\,
\ln\Biggl[\frac{(\vec b-\vec\rho/2)^2}
{\vec b+\vec\rho/2)^2}\Biggr] =
Z\alpha\,\ln\Biggl(\frac{b^2- b\rho\cos \phi + \rho^2/4}
{b^2 + b\rho\cos \phi + \rho^2/4}\Biggr)\ .
\label{b160}
 \eeqn
Substituting this expression in (\ref{b80}) we get $\sigma_1(\rho)$ in the 
following form,
 \beq
\sigma_1(\rho) = 2\,\int\limits_0^{b_0}db\,b
\int\limits_0^{2\pi} d\phi\,\left[ 1- 
\left(\frac{b^2- b\rho\cos \phi + \rho^2/4}
{b^2 + b\rho\cos \phi + \rho^2/4}
\right)^{iZ\alpha}\right]\ .
\label{b170}
 \eeq

To proceed to the further modifications of this expression we employ the
relation,
 \beq
1-X^Y = - XY\,F(1-Y,1;2;1-X)\ ,
\label{b180}
 \eeq
with
 \beqn
X &=& \frac{b^2- b\rho\cos \phi + \rho^2/4}
{b^2 + b\rho\cos \phi + \rho^2/4}\ ,\nonumber\\
Y &=& iZ\alpha\ .
\label{b190}
 \eeqn
 We also make use of the following representation for the hypergeometric
function,
 \beq
F(1-Y,1;2;1-X) = 
\frac{1}{\Gamma(1-Y)\,\Gamma(1+Y)}
\int\limits_0^1 dt\ 
\frac{t^{-Y}\,(1-t)^Y}{1-t(1-X)}\ .
\label{b200}
 \eeq
 Applying these relations to (\ref{b170}) we get,
 \beq
\sigma_1(\rho) = - \frac{2}
{\Gamma(1-iZ\alpha)\,\Gamma(1+iZ\alpha)}
\int\limits_0^1 dt\,t^{-iZ\alpha}\,
(1-t)^{iZ\alpha}\,L(b_0,\rho,t)\ ,
\label{b210}
 \eeq
 where
 \beqn
&& L(b_0,\rho,t) =
\int\limits_0^{b_0} db\,b
\int\limits_0^{2\pi} d\phi\ 
\frac{2\,b\,\rho\,\cos \phi}
{b^2+(1-2t)b\rho\cos \phi + \rho^2/4} =
2\pi\rho^2(1-2t)\nonumber\\
&\times& 
\left\{\frac{b_0^2}{b^2+\rho^2/4+\sqrt{D}} -
{1\over2}\,\ln\Biggl[
\frac{b^2+[1-2(1-2t)^2]\rho^2/4+\sqrt{D}}
{2\rho^2\,t(1-t)}\Biggr]\right\}\ ,
\label{b220}
 \eeqn
and 
 \beq
D=\left(b^2+{\rho^2\over4}\right)^2 -
b^2\rho^2\,(1-2t)^2\ .
\label{b225}
 \eeq

According to the above convention $\rho \ll b_0$ and this expression
can be further simplified,
 \beq
L(b_0,\rho,t) = 
\pi\,\rho^2\,(1-2t)\,\left\{1 - 
\ln\Biggl[\frac{b_0^2}{\rho^2\,t(1-t)}\Biggr]
+ O\Biggl(\frac{\rho^2}{b_0^2}\Biggr)\right\}\ .
\label{b230}
 \eeq

Now integration in (\ref{b210}) can be performed analytically, and
we arrive at the final expression for $\sigma_1$,
 \beq
\sigma_1(\rho) = 4\pi(Z\alpha)^2\,\rho^2\,
\left[\ln\Biggl(\frac{b_0}{\rho}\Biggr) -
2 - {\rm Re}\,\Psi(1+iZ\alpha) - C\right]\ ,
\label{b240}
 \eeq
where
 \beqn
\Psi(w) &=& \frac{d}{d\,w}\,\ln\Gamma(w)\ ;
\label{b250}\\
C &=& -\Psi(1) = 0.5772\ .
\nonumber
 \eeqn
 
Adding Eqs.~(\ref{b130}) and (\ref{b250}) we eventually get the dipole
cross section (\ref{b110}) in the form,
 \beq
\sigma(\rho) = \sigma_{Born}(\rho) - 
\Delta\sigma(\rho)\ .
\label{b260}
 \eeq
 Here
 \beqn
\sigma_{Born}(\rho) &=& 
4\pi(Z\alpha)^2\rho^2\,\left[
\ln\Biggl({a\over\rho}\Biggr) - 2\right];
\label{b270}\\
\Delta\sigma(\rho) &=& 
4\pi(Z\alpha)^2\rho^2\,\Bigl[ 
{\rm Re\,}\Psi(1+iZ\alpha) + C\Bigr] =
4\pi(Z\alpha)^2\rho^2\,f(Z\alpha)\ ,
\label{b280}
 \eeqn
 and
 \beq
f(Z\alpha) = (Z\alpha)^2\,
\sum\limits_{k=1}^\infty\ 
\frac{1}{k[k^2+(Z\alpha)^2]}\ .
\label{b290}                             
 \eeq

Now we are in position to calculate the photoproduction cross section substituting 
Eq.~(\ref{b260}) into (\ref{10}),
 \beq
\sigma(\gamma Z\to e^+e^-Z) = 
\sigma_{Born}(\gamma Z\to e^+e^-Z) - 
\Delta\sigma(\gamma Z\to e^+e^-Z)\ ,
\label{b300}
 \eeq
where
 \beqn
\sigma_{Born}(\gamma Z\to e^+e^-Z) &=&
\frac{28\,Z^2\alpha^3}{9\,m^2}\ 
\left[\ln(2am)-C-{83\over42}\right]\ ;
\label{b310}\\
\Delta\sigma(\gamma Z\to e^+e^-Z) &=&
\frac{28\,Z^2\alpha^3}{9\,m^2}\ f(Z\alpha)\ .
\label{b320}
 \eeqn

Note that the Coulomb correction $\Delta\sigma(\gamma Z\to e^+e^-Z)$ in
Eq.~(\ref{b320})  coincides with one calculated in \cite{dbm} using a
very different technique. At the same time, the Born terms are different
what one should have expected. Indeed, the Bethe-Maximon theory describes
photoproduction of $e^+e^-$ pairs off a point like nonscreened nucleus,
while we are dealing with pair production off atoms. Equivalence of the
Coulomb corrections calculated within the eikonal approximation and in
the Bethe-Maximon theory in the case of photon bremsstrahlung can be
proven exactly in the same way, since both processes are controlled by
the same dipole cross section.

\end{document}